\documentclass[aps,a4paper,twocolumn,prl]{revtex4-1}
\usepackage{amsmath}%
\usepackage{amsfonts}%
\usepackage{amssymb}%
\usepackage[cal=cm]{mathalfa}
\usepackage{graphicx}
\usepackage{braket}
\usepackage{sidecap}
\usepackage{color}
\usepackage{bbm}
\usepackage{nicefrac}
\usepackage{float}
\usepackage[dvipsnames]{xcolor}
\usepackage[utf8]{inputenc}
\usepackage[normalem]{ulem}
\usepackage{soul}
\usepackage{nicefrac}
\usepackage{sidecap}
\usepackage{xcolor, soul} 
\usepackage{physics}
\usepackage{amsmath,amssymb,amsthm,mathrsfs,amsfonts,dsfont} 
\usepackage{braket}
\usepackage{enumerate}
\usepackage{fancyhdr}
\usepackage{fancyhdr}
\usepackage{ulem}
\newcommand{\cyan}{\textcolor{black}} %
\newcommand{\blue}{} %
\newcommand{\red}{\textcolor{black}} %
\newcommand{\cut}[1]{}

\begin{document}

\title{Neutron optical test of completeness of quantum root-mean-square errors}

\author{Stephan Sponar$^{1}$}
\email{stephan.sponar@tuwien.ac.at}
\author{Armin Danner$^{1}$}
\author{\cyan{Masanao Ozawa}$^{2,3}$}
\author{Yuji Hasegawa$^{1,4}$}

\affiliation{%
$^1$Atominstitut, TU Wien, Stadionallee 2, 1020 Vienna, Austria \\
$^2$Graduate School of Informatics, Nagoya University, Chikusa-ku, Nagoya 464-8601, Japan\\
$^3$College of Engineering, Chubu University,1200 Matsumoto-cho, Kasugai 487-8501, Japan\\
$^4$Department of Applied Physics, Hokkaido University, Kita-ku, Sapporo 060-8628, Japan}

\begin{abstract}
\begin{center}
\today
\end{center}
%Defining and quantifying the error of a measurement is fundamental in any experimental science. 
%However, quantum physics shows a peculiar difficulty.
%One of the major problems, which has eluded a satisfactory solution for a long time, % so far,
%has been to generalize the notion of the classical root-mean-square error to quantum measurements
%to obtain an error measure satisfying both soundness and completeness, where
%an error measure is called sound if it vanishes for any accurate measurements, and called complete if
%it vanishes only for accurate measurements.
%The noise-operator based error measure has been commonly used as a sound error measure
%generalizing the classical root-mean-square error, but Busch, Heinonen, and Lahti found a case where
%it shows incompleteness.
%Recently, Ozawa proposed a modified definition for a noise-operator based error measure to overcome
%the incompleteness to obtain a sound and complete generalization of the classical root-mean-square error
%in \emph{npj Quantum Inf.~{\bf 5}, 1 (2019)}. Here, we present a neutron optical demonstration for the completeness of the new error measure for both projective (or sharp) measurements 
%as well as generalized (or unsharp) measurements described by positive-operator-valued measures.
%
%(short)
\blue{One of the major problems in quantum physics has been to generalize the classical root-mean-square error to quantum measurements to obtain an error measure satisfying both soundness (to vanish for any accurate measurements) and completeness (to vanish only for accurate measurements).
A noise-operator based error measure has been commonly used for this purpose, but it has turned out incomplete.}  Recently, Ozawa proposed a \blue{new definition for a noise-operator based error measure to be both sound and complete.} Here, we present a neutron optical demonstration for the completeness of the new error measure for both projective (or sharp) as well as generalized (or unsharp) measurements.
\end{abstract}

\maketitle
\section{ Introduction}

The notion of the mean error of any measurement is a well-defined quantity in classical physics. However, extending the classical notion of root-mean-square (rms) error, which has been broadly accepted as the standard definition for the mean error of a measurement,  to quantum measurements is a highly challenging and a non-trivial task \cite{Ozawa03,Ozawa04,Hall04,Branciard13,Busch13,Busch13PRA,Busch14,Buscemi14}. 
A straightforward generalization is represented by the \emph{noise-operator} based quantum root-mean-square (q-rms) error, \blue{defined as the root-mean-square of the noise operator},
where the noise-operator indicates how closely a meter observable ``tracks'' the observable \blue{to be} measured.
 \blue{The noise-operator} was first used in attempts to prove Heisenberg's uncertainty relation \cite{Heisenberg27}  for \blue{approximate} simultaneous measurements of pairs of non-commuting \blue{observables \cite{Arthurs65,Yamamoto86,Arthurs88}.}
%\cut{arbitrary, as well as, conjugate observables} 
%by Arthurs--Kelly \cite{Arthurs65},
%Yanamoto--Haus \cite{Yamamoto86}, and 
%. The noise operator was later utilized to formulate a generalized Heisenberg uncertainty relations for simultaneous measurements, when the measuring apparatus is considered as well by
%Arthurs--Goodman \cite{Arthurs88}.}

In more recent developments the noise-operator \blue{based q-rms error} was used to reformulate Heisenberg's \blue{error-disturbance relation} to be universally valid \cite{Ozawa03,Ozawa04} and made the conventional formulation testable. 
\blue{The validity of the new formulation as well as the violation of the conventional formulation 
were observed} first in neutronic \cite{Erhart12, Sulyok13, Demirel16,Sponar17,Demirel19,Demirel20} and in photonic  \cite{Steinberg12,Edamatsu13,Kaneda14,Ringbauer14,Ma16,Pan19} systems for successive spin measurements. 

However, Busch, Heinonen, and Lahti (BHL) \cite{Busch04} raised a completeness problem for the noise-operator based quantum root-mean-square error, which eluded a solution for a decade and brought about a debate on the
use of the noise-operator \cite{Busch14}, until Ozawa \cite{Ozawa19} recently brought a satisfactory solution. It is the purpose of this work to briefly recapitulate the argument of \blue{BHL} \cite{Busch04}, Ozawa's 
\blue{new definition of a sound and complete noise-operator based q-rms error \cite{Ozawa19}}, and to present a neutron optical experiment which demonstrates the \blue{completenesss} of the \blue{new} noise-operator based \emph{q-rms error}. At this point we want to emphasize that the new error notion maintains the previously obtained universally valid uncertainty relations and their experimental confirmations without changing their forms and interpretations \cyan{\cite{Ozawa19}}. 
\begin{figure}[!b]
\begin{center}
	\includegraphics[width=0.5\textwidth]{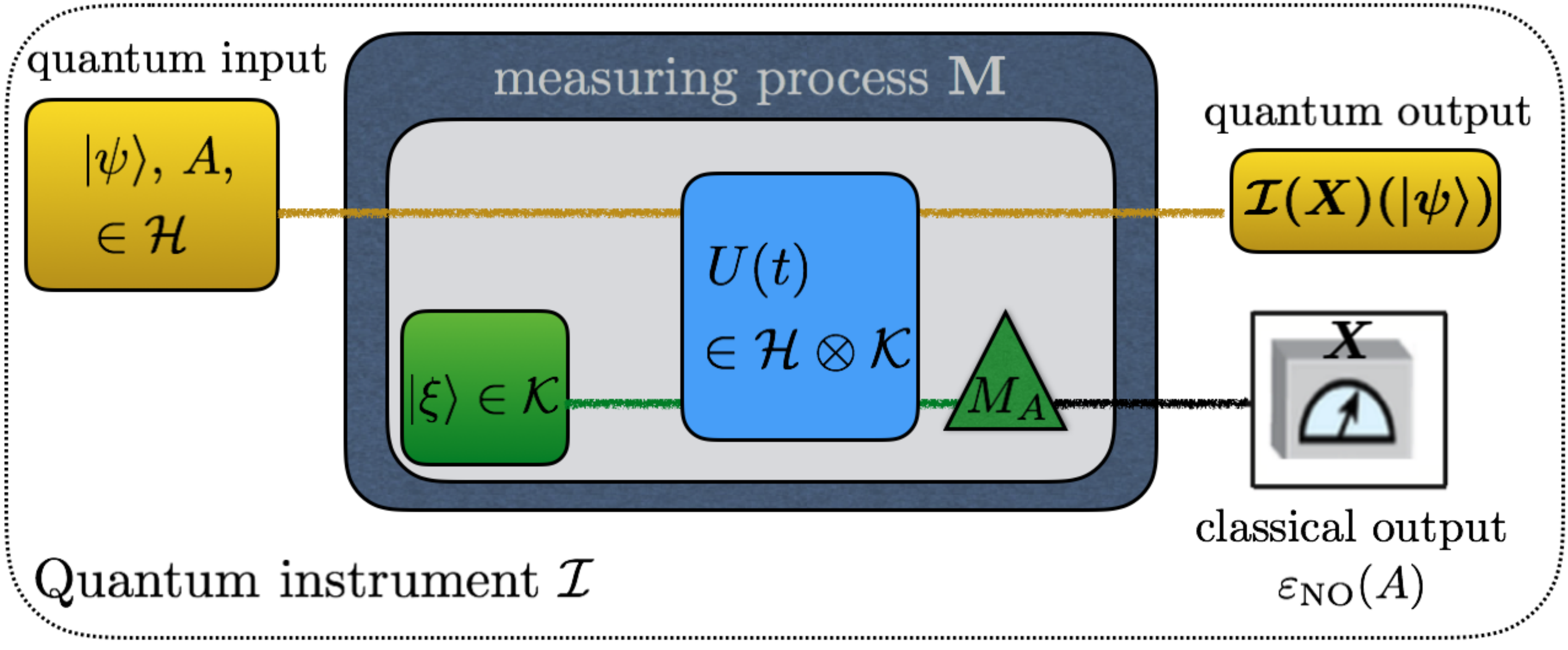}
	\caption{Schematic illustration of the \emph{indirect measurement model} with noise-operator based quantum root-mean-square (q-rms) error $\varepsilon_{\textrm{NO}}(A)$ of measuring process $\textbf{M}$.}
	\label{fig:model}
\end{center}
\end{figure}
\vspace{-3mm}

\section{Theory}

The noise-operator based quantum root-mean-square (q-rms) error \cite{Ozawa05} of a measuring process $\textbf{M}$, on quantum instrument $\mathcal{I}$, is denoted as $\varepsilon_{\textrm{NO}}(A)=\langle\psi,\xi\vert N(A,\textbf{M})^2\vert\red{\xi,\psi}\rangle^{1/2}$, where the \emph{noise-operator} $N(A,\textbf{M})$  describes how accurately the value of an observable $A$ is transferred to the meter observable $M_A$, during the evolution $U(t)$ of the composite system: $N(A,\textbf{M})= U(t)^\dagger\cyan{( {1\!\!1} \otimes M_A)}U(t)- A\otimes {1\!\!1} $. Here $A$ is an observable of a system $\textbf{S}$ in state $\vert\psi\rangle$ of Hilbert space $\mathcal H$, and $M_A$ is the observable representing the meter of the observer in the probe system (measurement device) $\textbf{P}$ in initial state $\vert \xi\rangle$ of Hilbert space $\mathcal K$. Moreover, $U(t)$ is the unitary evolution of the composite quantum system $\textbf{S}+\textbf{P}$. This concept, introduced in \cite{Ozawa84}, is usually referred to as \emph {indirect measurement model} \red{of measuring process $\textbf{M}$} and schematically illustrated in Fig.\,\ref{fig:model}.

\vspace{3mm}

In the Heisenberg picture, we shall write $A(0)=A\otimes 1\!\!1$ and $M_A(t)=
U(t)^{\dagger}({1\!\!1}\otimes M_A)U(t)$.
The POVM $\Pi$ of the measuring process $\textbf{M}$ is defined by
$\Pi(x)=\langle \xi\vert {\rm P}^{M_A(t)}(x)\vert\xi\rangle$,
where ${\rm P}^{M_A(t)}(x)$ is the spectral projection of $M_A(t)$ for eigenvalues $x$.
The {\em moment operator} $M$ of the POVM $\Pi$ is defined by $M=\sum_{x}x\,\Pi(x)$,
and the {\em second moment operator} $M^{(2)}$ of the POVM  $\Pi$ is defined by
$M^{(2)}=\sum_{x}x^2\,\Pi(x)$.
The measurement is called a {\em sharp measurement of $M$} if $\Pi$ is projection-valued.
In this case, we have $\Pi(x)=P^{M}(x)$ and $M^{(2)}=M^2$. 
Otherwise, the measurement is called an {\em unsharp (or generalized) measurement of $M$};
in this case we have $M^{(2)}> M^{2}$.
An important property of $\varepsilon_{\textrm{NO}}(A)$ is that it is determined by
(moment operators of)
the POVM $\Pi$ of \textbf{M} in such a way that
\begin{equation}\label{eq:errordef}
\varepsilon_{\textrm{NO}}(A,\Pi,\vert\psi\rangle)^2=
\langle\psi\vert (A-M)^2\vert\psi\rangle+\langle\psi\vert(M^{(2)}-M^2)\vert\psi\rangle.
\end{equation}
\red{This property and its consequence are to be studied in the present experiment. }

\subsection {Counter-example}\label{sec:counter}

\blue{It is shown by BHL \cite{Busch04}} that \blue{there exists} a measuring process $\textbf{M}$ with $\varepsilon_{\textrm{NO}}(A,\Pi,\vert\psi\rangle)=0$, 
whereas $\textbf{M}$ does not accurately measure $A$. However, a vanishing error is only expected for an accurate measurement for the completeness of the error measure.
Here, we do not give the original counter-example but the slightly simplified version as stated in \cite{Ozawa19}.
Consider a measurement of the observable $A$ in a two-level system in the initial state $\ket{\psi}$
with measuring process described by a POVM $\Pi$ with the moment operator $M$ 
given as follows.
\begin{eqnarray}\label{eq:example}
A=\left( \begin{array}{rrrr}
1 & 1   \\
1 & 1  \\
\end{array}\right),\quad M=\left( \begin{array}{rrrr}
1 & 1   \\
1 & -1  \\
\end{array}\right),\quad \vert\psi\rangle=\left( \begin{array}{rrrr}
1    \\
0   \\
\end{array}\right).
\end{eqnarray}

First, we consider the sharp measurement of $M$ with the POVM $\Pi_1$.
In this case, one obtains 
%Then \blue{for the measuring process $\textbf{M}$ such that 
$\Pi_1(x)={\rm P}^{M}(x)$ and $M^{(2)}=M^2$, so that 
%one obtains 
%
\begin{eqnarray}\label{eq:examperesult}
\varepsilon_{\textrm{NO}}(A,\Pi_1,\vert\psi\rangle)&=&\langle\psi\vert (A-M)^2\vert\psi\rangle^{1/2}=0.
%,
\end{eqnarray}
%
%with  $\langle\psi\vert A\vert\psi\rangle=\langle\psi\vert M\vert\psi\rangle=1$.
However, this particular measurement is not accurate, since $A$ and $M$ have disjoint spectra.
%spectrum. 
The operator $A$ has spectral decomposition $A=\sum_i  a_i\ketbra{a_i}$, with eigenvalues $a_i=\{2,0\}$ and normalized eigenvectors $\vert a_i\rangle=1/\sqrt{2}(1,\pm1)^T=\ket {\pm x}$, while $M=\sum_i m_i \ketbra{ m_i}$, with eigenvalues  $m_i=\{\pm\sqrt{2}\}$ and normalized eigenvectors $\vert m_i\rangle=\{\frac{1}{\sqrt{4+2\sqrt 2}}(1+\sqrt{2}, 1)^T,\frac{1}{\sqrt{4-2\sqrt 2}}(1-\sqrt{2}, 1)^T\}$. With $ {\rm{P}}^A(2)$, being the projector associated with eigenvalue 2, that is $\ketbra{+x}\equiv  {\rm{P}}^{\sigma_x} (1) $, which finally gives $\bra{+z}  {\rm{P}}^{\sigma_x} (1) \ket{+z}=\frac{1}{2}$. We can then write $\langle\psi\vert {\rm{P}}^A(2)\vert\psi\rangle=\frac{1}{2}\neq\langle\psi\vert \Pi_1(2)\vert\psi\rangle=0$ to express the inaccuracy of the measurement. Thus, the measurement with the POVM $\Pi_1=P^{M}$ does not accurately measure $A$ but  $\varepsilon_{\textrm{NO}}(A,\Pi_1,\vert\psi\rangle)=0$.

Secondly, we consider the unsharp measurement of $M$ with POVM $\Pi_2$ given by
\begin{eqnarray}  \label{eq:POVM}
\Pi_2(2)&=&\frac{1}{2}\bigg( {1\!\!1} +\frac{1}{2} \sigma_x+\frac{1}{2} \sigma_z\bigg),\nonumber\\
\Pi_2(-2)&=&\frac{1}{2}\bigg( {1\!\!1} -\frac{1}{2}\sigma_x-\frac{1}{2} \sigma_z\bigg),
\end{eqnarray}
for which we have $M=2\Pi_2(2)-2\Pi_2(-2)$, so that the POVM $\Pi_2$ is an
unsharp measurement of $M$.  Then, we have $M^{(2)}=4{1\!\!1}$ and 
 $M^{2}=2{1\!\!1}$.  Thus, we have
$\varepsilon_{\textrm{NO}}(A,\Pi_2,\vert\psi\rangle)=\sqrt{2}.$
Since 
\[\varepsilon_{\textrm{NO}}(A,\Pi_2,\vert\psi\rangle)=
(\varepsilon_{\textrm{NO}}(A,\Pi_1,\vert\psi\rangle)^2+2)^{1/2}=\sqrt{2},
\]
the value $\varepsilon_{\textrm{NO}}(A,\Pi_2,\vert\psi\rangle)=\sqrt{2}$ will be revised when
the value $\varepsilon_{\textrm{NO}}(A,\Pi_1,\vert\psi\rangle)=0$ is revised for the completeness of
the error measure.

\subsection {Requirements}
To resolve this inconsistency, Ozawa introduced four requirements for a valid definition of error measure $\varepsilon$ generalizing the classical rms error \cite{Ozawa19}: 
%
%\begin{itemize}
\begin{enumerate}[(i)]
\item \emph{Operational definability}: The error measure is definable by 
he POVM $\Pi$ of measuring process $\textbf{M}$ with $A$ and $\vert\psi\rangle$,
i.e., $\varepsilon=\varepsilon(A,\Pi,\vert\psi\rangle)$.

\item \emph{Correspondence principle}: 
If $A(0)$ and $M_A(t)$ commute, 
$\varepsilon(A,\Pi,\vert\psi\rangle)$ equals the classical rms error determined 
by the joint probability distribution of $A(0)$ and $M(t)$.
\item \emph{Soundness}: The error measure $\varepsilon$ should vanish for any accurate measurements.
\item \emph{Completeness}: The  
converse of soundness - a measurement should be accurate if the error measure $\varepsilon$ vanishes.
\end{enumerate}
%\end{itemize}
%

Ozawa \cite{Ozawa19} showed that the noise-operator-based quantum rms error satisfies
requirements (i) -- (iii) so that it is a sound generalization of the classical rms error,
and proposed a modification of it to satisfy all the requirements (i) -- (iv) including completeness.
In addition to (i) -- (iv), the new error measure is shown to have the following two properties:
\begin{enumerate}[(i)]
	\setcounter{enumi}{4}
\item \emph{Dominating property}: The error measure $\varepsilon$ dominates the noise-operator based q-rms error, that is  $\varepsilon_{\textrm{NO}}(A,\Pi,\vert\psi\rangle)\leq \varepsilon (A,\Pi,\vert\psi\rangle)$.
\item \emph{Conservation property} for dichotomic measurements: The error measure $\varepsilon$ coincides with the noise-operator based q-rms error $\varepsilon_{\textrm{NO}}$ for dichotomic measurements, i.e., 
$\varepsilon_{\textrm{NO}}(A,\Pi,\vert\psi\rangle)=\varepsilon (A,\Pi,\vert\psi\rangle)$  if  
$A^2=M^{(2)}= {1\!\!1}$.
\end{enumerate}

Thus the new notion maintains all previously obtained universally valid uncertainty relations and their experimental confirmations \cite{Demirel20,Demirel19,Sponar17,Demirel16,Sulyok13,Erhart12} without changing their forms and interpretations, in contrast to a prevailing view that a state-dependent formulation for measurement uncertainty relation is not tenable \cite{Busch14}.

\subsection{Definition \& predictions of locally uniform quantum root-mean-square error}

For any $t \in  \mathbb{R}$ the \emph{quantum root-mean-square (q-rms) error profile} $\varepsilon_t$ for $A$ and $\Pi$ in $\vert\psi\rangle$ is defined as 
\begin{equation}\label{eq:errorprofile}
\varepsilon_t (A,\Pi,\vert\psi(t)\rangle) = \varepsilon_{\textrm{NO}}(A,\Pi, e^{- \mathrm{i} t A}\vert\psi\rangle). 
\end{equation}  
In order to obtain a numerical error measure the \emph{locally uniform q-rms error}  $\bar\varepsilon$ is given by 
\begin{equation}\label{eq:erroruniform}
\bar\varepsilon(A,\Pi\,\vert\psi\rangle)=\sup_{t\in  \mathbb{R}} \varepsilon_t (A,\Pi,\vert\psi(t)\rangle).
\end{equation}  
Then $\bar\varepsilon$ is a sound and complete q-rms error, satisfying both the dominating property, and the conservation property for dichotomic measurements. 

For the given example \red{from Eq.(\ref{eq:example})}, with $A,\,\Pi_1$, and $\ket{\psi}$, we get  
\begin{equation}\label{eq:error}
\varepsilon_t (A,\Pi_1,\vert\psi(t)\rangle) = 2\vert\sin t\vert \textrm{ and }
\bar\varepsilon (A,\Pi_1,\vert\psi\rangle) =2,
\end{equation}
for the sharp $M$ measurement described by the POVM $\Pi_1$.
The relation $\bar\varepsilon (A,\Pi,\vert\psi\rangle) =2$ correctly indicates that the measurement of $A$ described in the example above is not an accurate measurement. 
For the unsharp $M$ measurement described by the POVM $\Pi_2$, one gets
\begin{equation}\label{eq:errorPOVM}
\varepsilon_t (A,\Pi_2,\vert\psi(t)\rangle) =\sqrt{4-2\cos (2t)}\textrm{ and }\bar\varepsilon (A,\Pi_2,\vert\psi\rangle) =\sqrt{6}.
\end{equation}
Thus, the value $\epsilon_{\rm{NO}}(A,\Pi_2,\ket{\psi})=\sqrt 2$ is revised as  $\bar\epsilon(A,\Pi_2,\ket{\psi})\,=\,\sqrt 6$ for the completeness of the error measure $\bar\epsilon$.
\section{Experimental}

\begin{figure}[t]
	\includegraphics[width=0.5\textwidth]{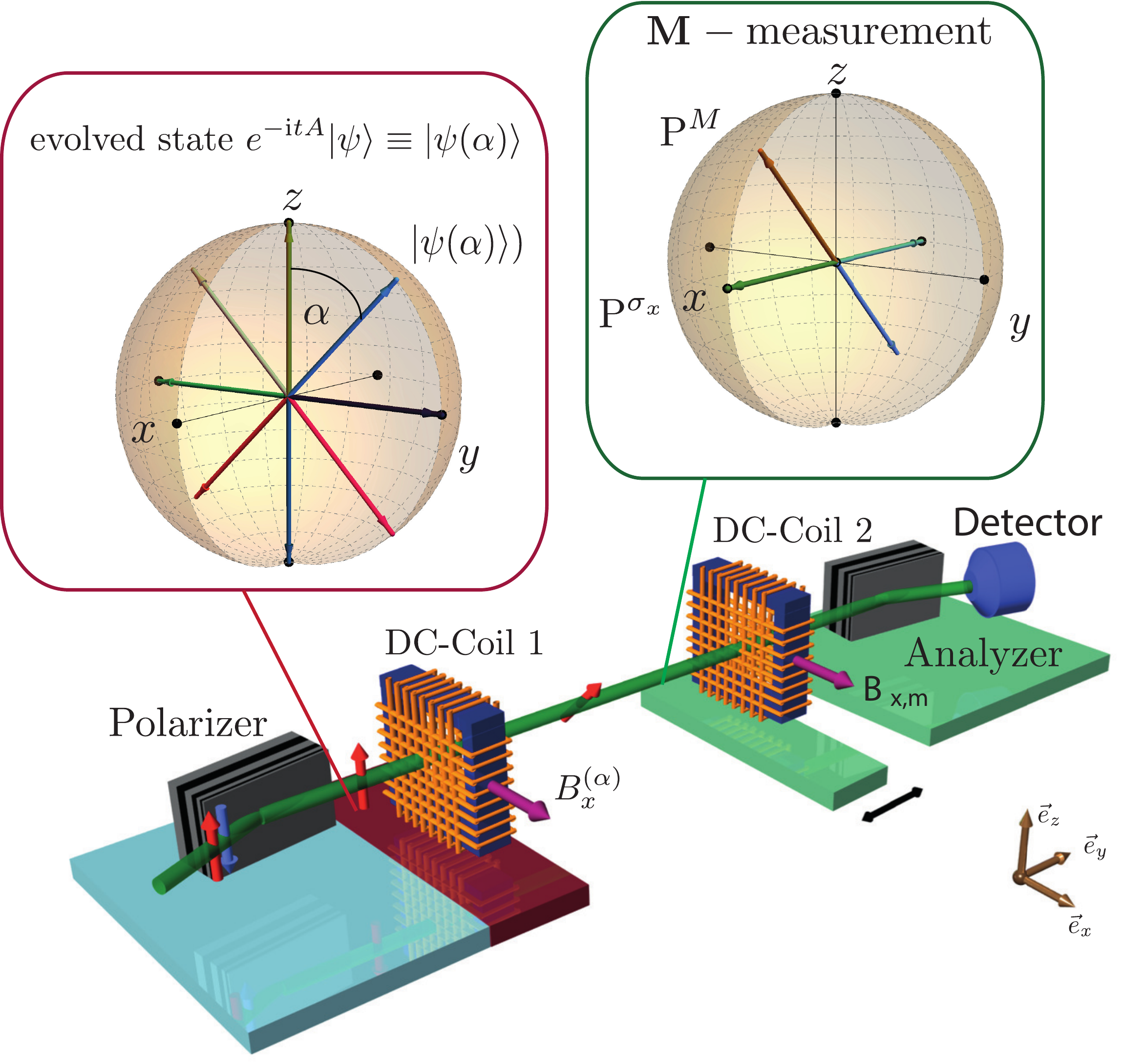}
	\caption{Experimental setup for measurement of the \emph{q-rms error profile} $\varepsilon_{\alpha}$. The setup consists of three regions:  Blue: preparation of the initial state $\vert\psi\rangle=(1,0)^T\equiv\ket{+z}$.  Red: preparation of the evolved state $ \vert\psi(t)\rangle=e^{- \mathrm{i} t A}\vert\psi\rangle \rightarrow\, \vert\psi(\alpha)\rangle=e^{- \mathrm{i} t \sigma_x}\vert\psi\rangle=e^{( \mathrm{i}\alpha  \sigma_x)/2}\vert\psi\rangle$. Green: Measurement of $A^2, M^2$ and $M$ in state $\vert \psi(\alpha)\rangle$, $\,\vert \psi(\alpha+\pi)\rangle$ and  $\ket{ +x}$, respectively. Projective (sharp) measurement are realized by applying projectors  $P^{M}(\pm \sqrt{2})$ and generalized (unsharp) in terms of POVM by randomized sequences of ${\rm P}^M$ and  ${\rm P}^{\sigma_x}$. \red{Bloch spheres above setup indicate evolution of initial state $\ket{\psi(\alpha)}$ and measured projectors ${\rm P}^M$ and  ${\rm P}^{\sigma_x}$.} }
	\label{fig:setup}
\end{figure}

Here, we present a neutron polarimetric measurement  of the \emph{quantum root-mean-square (q-rms) error profile} $\varepsilon_{\red{\alpha}}$, \red{ resulting in determination of the \emph{locally uniform q-rms error} $\bar{\epsilon}$}, for the POVM $\Pi_1$ (the sharp measurement of $M$) and the POVM $\Pi_2$ (an unsharp measurement of $M$), as given in Eqs.\,(\ref{eq:error}), (\ref{eq:errorPOVM}) to demonstrate the completeness property and thereby confirm the resolution of the inconsistency in question. 

\vspace{-3mm}

\subsection{Experimental setup}
The experiment was performed at the polarimeter instrument \emph{NepTUn (NEutron Polarimeter TU wieN)}, located at the tangential beam port of the 250\,kW TRIGA Mark II  research reactor at the Atominstitut - TU Wien, in Vienna, Austria. A schematic illustration of the setup is given in Fig.\,\ref{fig:setup}.  An incoming monochromatic neutron beam, reflected from a pyrolytic graphite crystal, with mean wavelength $\lambda\simeq 2.02\,\AA$ ($\Delta\lambda/\lambda\simeq0.02$) is polarized along the vertical  ($+z$) direction by refraction from a CoTi multilayer array, hence on referred to as supermirror.
The neutron polarimetric setup consists of three stages, as indicated in Fig.\,\ref{fig:setup}. The blue stage indicates the preparation of the incident state $\vert \psi\rangle=\ket{ +z}$, which is reflected from the polarizer (first super mirror). In the red stage the state evolution of initial state $\vert \psi\rangle=\ket{+z}$ as $\vert\psi(t)\rangle=e^{-{\rm{i}}t  A} \vert \psi\rangle\,\rightarrow\, e^{( \mathrm{i}\alpha  \sigma_x)/2}\vert\psi\rangle\equiv\vert\psi(\alpha)\rangle$ is induced, due to rotation by angle $\alpha$ about the $x$-axis \red{(note that the error profile $\epsilon_\alpha$ is a function of the rotation angle $\alpha$)}. The Larmor precession inside direct current (DC) coil 1 is induced by the static magnetic field $B_x^{(\alpha)}$.

In the green stage, a \emph{projective} (or \emph{sharp}) measurement of $M$ is performed first, in order to demonstrate the counter example $\varepsilon_t (A,\Pi_1,\vert\psi(t)\rangle) =2\vert\sin t\vert$ from \cite{Ozawa19}. The $\Pi_1$ measurement  has two possible outcomes, namely $m=+\sqrt 2$ and $m=-\sqrt 2$, corresponding to measurement operators $\Pi_1(\pm\sqrt 2)={\rm P}^M(\pm\sqrt 2)=\frac{1}{2}({1\!\!1} \pm\sigma_m)$, with $\sigma_m=\frac{1}{\sqrt 2}\sigma_z+\frac{1}{\sqrt 2}\sigma_x$. The combined action of DC-coil 2 and the analyzer (second super mirror) realizes the respective projector.

\begin{figure}[!t]
\begin{center}
	\includegraphics[width=0.46\textwidth]{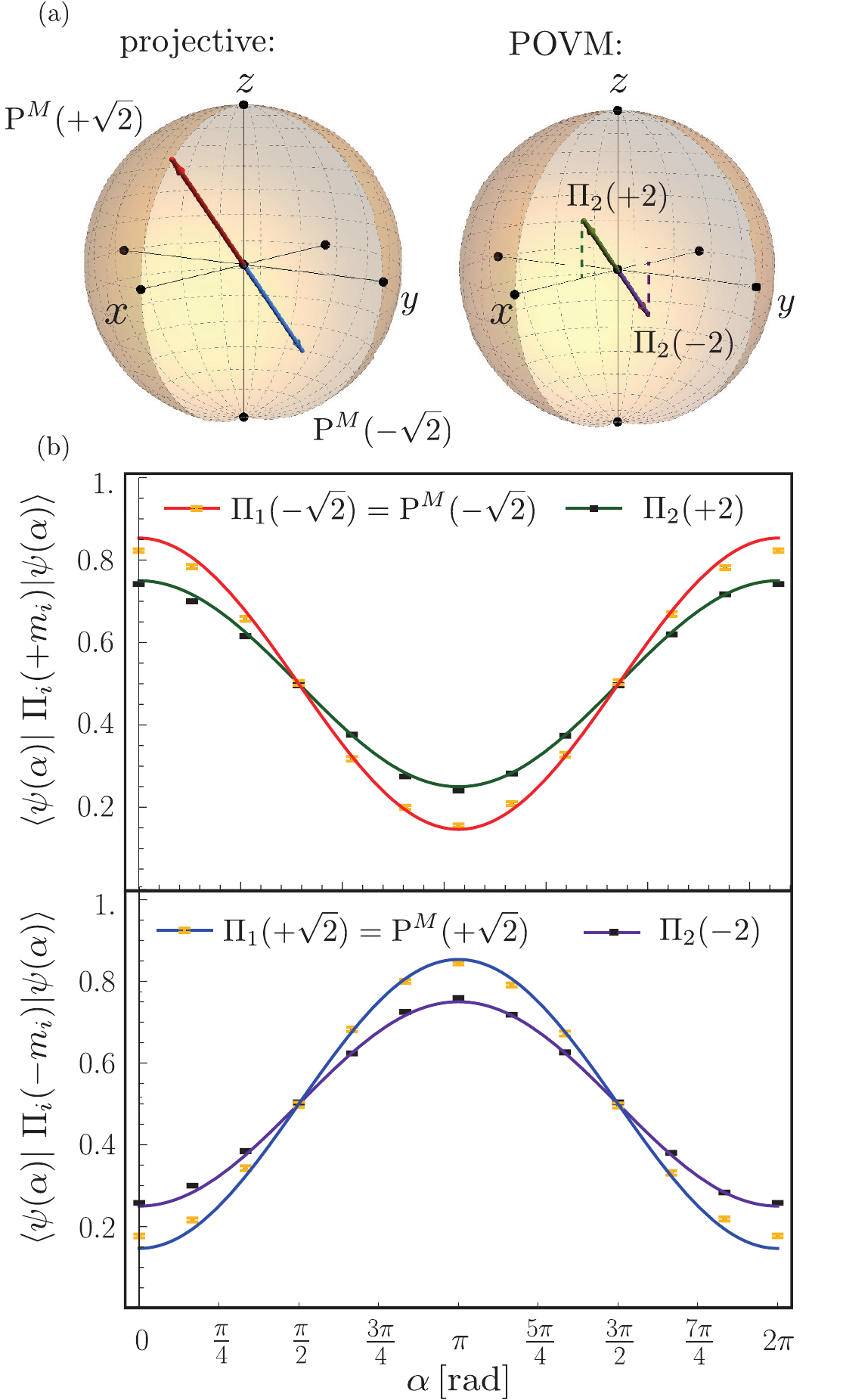}
	\caption{\textcolor{black}{(a) Bloch sphere depiction of projectors $\Pi_1(\pm\sqrt 2)={\rm P}^M(\pm\sqrt 2)$ and POVM elements $\Pi_2(\pm 2)$. (b) Expectations values of $\bra{\psi(\alpha)} \Pi_1(\pm\sqrt 2) \ket{\psi(\alpha)}=\bra{\psi(\alpha)} {\rm P}^M(\pm\sqrt 2) \ket{\psi(\alpha)}$  for projective and $\bra{\psi(\alpha)} \Pi_2(\pm 2) \ket{\psi(\alpha)}$  for generalized POVM in state $\ket{\psi(\alpha)}$ for measurement time $t_{\rm{maes}}=100\,$sec. (error-bars represent $\pm 1$ st. dev.).}}
	\label{fig:pr}
	\end{center}
\end{figure}

In addition to the sharp measurement, we also \red{realized} a \emph{generalized} (or \emph{unsharp}) measurement of $M$ in terms of a positive-operator-valued measures (POVM) elements $\Pi_2(\pm 2)=\frac{1}{2}( {1\!\!1} \pm\frac{1}{2} \sigma_x\pm\frac{1}{2} \sigma_z)$, yielding q-rms error profile $\varepsilon_\alpha (A,\Pi_2,\vert\psi(\alpha)\rangle)$. This is \red{achieved} by a randomized combination of projectors of ${\rm P}^M (\pm\sqrt 2)$ together with a contribution of a \emph{no-measurement}, realized by the $\pm x$ projectors, denoted as ${\rm P}^{\sigma_x}(\pm 1)$, which gives ($\langle\psi(\alpha)\vert{\rm P}^{\sigma_x}(\pm 1) \vert\psi(\alpha)\rangle=\frac{1}{2}\,\,$ for all $\alpha\in [0,2\pi]$) (see Sec. Methods for experimental details of the POVM realization). 

\begin{figure}[!b]
\begin{center}
	\includegraphics[width=0.48\textwidth]{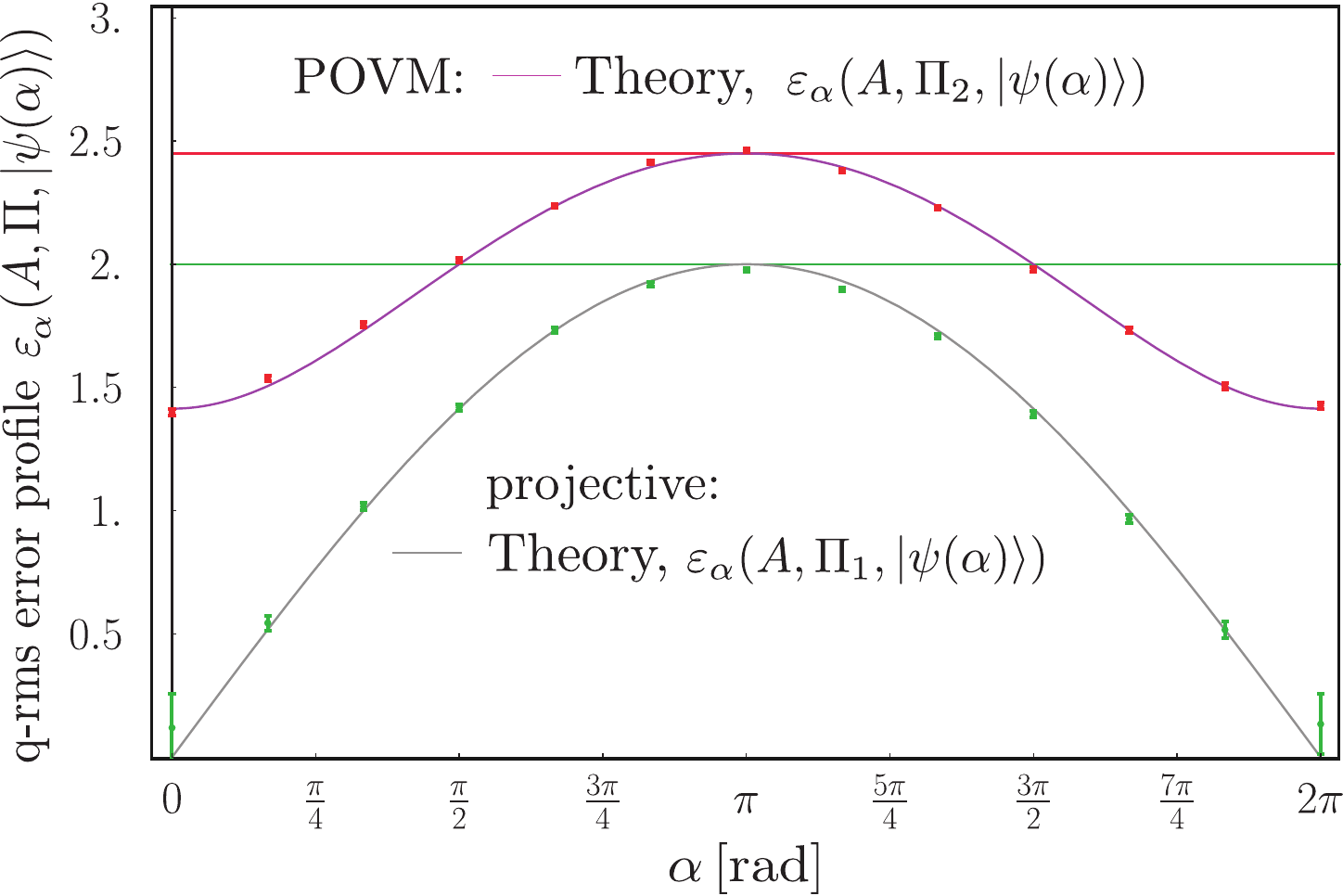}
	\caption{\textcolor{black}{Final experimental results of quantum root-mean-square (q-rms) error profile $\varepsilon_\alpha (A,\Pi_1,\vert\psi(\alpha)\rangle)$ (projective) and $\varepsilon_\alpha (A,\Pi_2,\vert\psi(\alpha)\rangle)$ (POVM)  of measurement process $\textbf M$, for different evolved states $\vert\psi(\alpha)\rangle$ for measurements time $t_{\rm{maes}}=100\,$sec (error-bars represent $\pm 1$ st. dev.). \emph{Locally uniform q-rms error} $\bar\varepsilon (A,\Pi_1,\vert\psi)=2$ and  $\bar\varepsilon (A,\Pi_2,\vert\psi)=\sqrt 6$} are represented by green and red line, respectively.}
	\label{fig:ep}
	\end{center}
\end{figure}

\red{Figure \ref{fig:pr}\,(a) gives a Bloch sphere depiction of projectors $\Pi_1(\pm\sqrt 2)={\rm P}^M(\pm\sqrt 2)$ and POVM elements $\Pi_2(\pm 2)$. The finally} recorded intensity was about 350 neutrons s$^{-1}$ at a beam cross-section of 10 (vertical) $\times$ 5 (horizontal) mm$^2$.  A  $^3$He detector with high efficiency (more than 99\,\%) is used for count rate detection. To avoid unwanted depolarization, a static guide field pointing in the $+z$-direction with a strength of about 10 Gauss is produced by rectangular Helmholtz coils. In addition, the guide field induces Larmor precession, which, together with two appropriately placed DC coils, enables state preparation of $\ket{\psi(\alpha)}$ and projective or generalized measurements $\Pi_1$ and $\Pi_2$. 

\subsection{Experimental results}

Experimental results of expectations values $\langle\psi(\alpha)\vert\Pi_i(\pm m_i)\vert\psi(\alpha\rangle)$ (with $i=1,2$ and $m_i=\{\sqrt2,2\}$), that is $\langle\psi(\alpha)\vert\Pi_1(\pm \sqrt2)\vert\psi(\alpha\rangle)=\langle\psi(\alpha)\vert  {\rm P}^{M}(\pm\sqrt 2) \vert\psi(\alpha\rangle)$ of projective (sharp) measurements and $\langle\psi(\alpha)\vert\Pi_2(\pm 2)\vert\psi(\alpha\rangle)$ of generalized (unsharp) POVM are plotted in Fig.\,\ref{fig:pr}\,(b). See Sec. Methods for details of the measurement procedure.  

The error-profile $\varepsilon_\alpha (A,\Pi_1,\vert\psi(\alpha)\rangle)$ is obtained by measuring expectation values of $A^2$, $M^2$ and $M$ in state $\vert\psi(\alpha)\rangle$ and of $M$ in auxiliary states  $\vert\psi(\alpha+\pi)\rangle$ and  $\ket{ +x}$ (see Sec. Methods for details of the experimental realization of sharp $M$ measurement process). 
The final results for the error profile $\varepsilon_\alpha (A,\Pi_1,\vert\psi(\alpha)\rangle)$ for projective $M$ measurement and $\varepsilon_\alpha (A,\Pi_2,\vert\psi(\alpha)\rangle)$ for generalized measurements (POVM), are plotted in Fig.\,\ref{fig:ep}. For the initial state $\ket{\psi}=\ket{+z}$, which corresponds to $\alpha=0$, the \emph{q-rms error profile} of the sharp (projective) measurement is zero; $\varepsilon_\alpha (A,\Pi_1,\vert\psi(\alpha=0)\rangle)=0$, as expected from the counter-example from Eq.(\ref{eq:examperesult}). The maximum value of  $\varepsilon_\alpha=2$ is obtained for $\alpha=\pi$, namely $\varepsilon_\alpha (A,\Pi_1,\vert\psi(\alpha=\pi)\rangle)=2$. From this we infer the value of the \emph{locally uniform q-rms error} $\bar\varepsilon(A,\Pi_1\,\vert\psi\rangle)=\sup\varepsilon_\alpha (A,\Pi_1,\vert\psi(\alpha)\rangle)$ as $\bar\varepsilon (A,\Pi_1,\vert\psi)=2$. As can be seen from Fig.\,\ref{fig:ep}, the theoretical prediction for the error profile $\varepsilon_\alpha (A,\Pi_1,\vert\psi(\alpha)\rangle)= 2\vert\sin \frac{\alpha}{2}\vert $ are evidently reproduced for all $\alpha\in[0,2\pi]$, as predicted in \cite{Ozawa19}. 

The generalized (unsharp) measurement in terms of POVMs  also reproduces the theoretical predictions of the q-rms error profile $\varepsilon_\alpha (A,\Pi_2,\vert\psi(\alpha)\rangle)= \sqrt{4-2\cos\alpha}$, with minimum value $\varepsilon_\alpha (A,\Pi_2,\vert\psi(\alpha=0,\,2\pi)\rangle) = \sqrt 2$ and maximum value $\varepsilon_\alpha (A,\Pi_2,\vert\psi(\alpha=\pi)\rangle) = \sqrt 6= \bar\varepsilon(A,\Pi_2,\vert\psi\rangle)$, the \emph{locally uniform q-rms error}. The higher values of the POVM error profile (meaning $\varepsilon_\alpha (A,\Pi_2,\vert\psi(\alpha)\rangle) > \varepsilon_\alpha (A,\Pi_1,\vert\psi(\alpha)\rangle)$ for all $\alpha\in[0,2\pi]$) are caused by the unsharp character of the POVM \red{measuring process.}

\section{Discussion  \& Conclusion}

As seen already from Eq.(\ref{eq:errordef}) the error $\varepsilon(A,\Pi,\vert\psi\rangle)$ depends on the choice of the respective POVM $\Pi$ that realizes a particular measurement. From a physical point of view one might ask which measurement is optimal\,? Although individual expectation values (mean values) are the same for sharp (projective) and unsharp (POVM) realizations of the same measurement $M$, regarding measurement error of single measurements, sharp measurements are always superior compared to unsharp measurements; the reason a clearly seen from Eq.(\ref{eq:errordef}). However in the case of joint,  \emph{simultaneous} (or \emph{successive}) measurements, where an optimal \emph{error-error} (or \emph{error-disturbance}) trade-off is required, unsharp measurements are able to outperform sharp measurements \cite{Branciard16,Demirel19,Sponar20POVM}. 

To conclude, despite numerous successful experimental demonstrations of error-disturbance uncertainty relations based on \blue{the noise-operator based q-rms error commonly used as a sound generalization
of the classical rms error \cite{Ozawa03,Ozawa04,Ozawa05,Erhart12,Steinberg12,Edamatsu13}},  
Busch, Heinonen, and Lahti \cite{Busch04} \blue{pointed out the incompleteness of this error measure.
% raised a problem %for quantum generalizations of the classical rms error by proposing a counter-example in \cite{Busch04} questioning the completeness of Ozawa's error definition.
A new definition for a noise-operator based error-measure remedying the incompleteness was recently 
proposed by Ozawa \cite{Ozawa19}.}
 It is important to note that the new error-notion affects only non-dichotomic measurements and consequently all experimentally obtained results up to date remain valid.
 %However, a  of the root-mean-square error was required, which was achieved by Ozawa in 2019 \cite{Ozawa19}.
The completeness behavior of the new root-mean-square definition of the error has been observed in detail along with the counter-example given in Ref.~\cite{Ozawa19}.

\section{Methods}

In order to experimentally demonstrate the completeness of locally uniform q-rms error $\bar\varepsilon$, Eqs.(\ref{eq:error}), (\ref{eq:errorPOVM}) need to be expressed in terms of experimentally accessible quantities, i.e., \emph{expectation values}. This can be achieved by applying the well known \emph{three-state-method} \cite{Ozawa05} for generalized measurements  \cite{Ozawa04} (p.383) to obtain the \emph{state-dependent q-rms error profile} $\varepsilon_t(A,\Pi,\vert\psi(t)\rangle)$ from
\begin{eqnarray}\label{eq:ErrPro}
\varepsilon^2_t(A,\Pi,\vert\psi(t)\rangle)&=&  \langle\psi(t)\vert (M-A)^2\vert\psi (t)\rangle  \nonumber\\ &+&  \langle\psi(t)\vert M^{(2)}-M^2\vert\psi(t)\rangle,
\end{eqnarray}  
where $M^{(2)}$ denotes the second moment of $\Pi$, given by $M^{(2)}=\sum_x x^2\, \Pi(x)$. 
%\begin{eqnarray}\label{eq:5stateNew}
%&&\varepsilon^2_t(A,\Pi,\vert\psi(t)\rangle)= \langle\psi(t)\vert M^{(2)}-M^2\vert\psi(t)\rangle\nonumber\\&+&\underbrace{\langle\psi(t)\vert A^2\vert\psi(t)\rangle}_{\neq 1}+\underbrace{\langle\psi(t)\vert M^{2}\vert\psi(t)\rangle}_{\neq 1}-{\langle\psi(t)\vert M\vert\psi(t)\rangle}\nonumber\\&-&\underbrace{\langle\psi(t)\vert A\, M\,A\vert\psi(t)\rangle}_{2\langle+x\vert M\vert+x\rangle}+\overbrace{\langle\psi(t)\vert \underbrace{(A-{1\!\!1})}_{\sigma_x}}^{\langle \psi^\bot(t)\vert}\, M\,\overbrace{\underbrace{(A-{1\!\!1})}_{\sigma_x}\vert\psi(t)\rangle}^{\vert\psi^\bot(t)\rangle}.\nonumber\\
%\end{eqnarray}  
 The first term of Eq.(\ref{eq:ErrPro}) can be symmetrized, applying the operator identity 
\begin{eqnarray}
\hspace{-5mm}(A-{1\!\!1})M(A-{1\!\!1})-A\,M\,A-M=-(M\, A+A\, M),\nonumber\\
\end{eqnarray}  
which gives 
\begin{eqnarray}\label{eq:3state}
\varepsilon^2_\alpha(A,\Pi,\vert\psi(\alpha)\rangle) &=&{\langle\psi(\alpha)\vert A^2\vert\psi(\alpha)\rangle}+{\langle\psi(\alpha)\vert M^{2}\vert\psi(\alpha)\rangle}\nonumber\\
&-&{\langle\psi(\alpha)\vert M\vert\psi(\alpha)\rangle}-{\langle\psi(\alpha)\vert A\, M\,A\vert\psi(\alpha)\rangle}\nonumber\\
&+&{\langle\psi(\alpha)\vert {(A-{1\!\!1})}}\, M\,{{(A-{1\!\!1})}\vert\psi(\alpha)\rangle}\nonumber\\&+& \langle\psi(\alpha)\vert M^{(2)}-M^2\vert\psi(\alpha)\rangle.
\end{eqnarray}  
The q-rms error-profile for all evolved states $\vert\psi(\alpha)\rangle)$ is calculated using the \emph{three-state method} (see Supplementary Information \cite{SuppPOVM} for the individual measurement results of all terms from Eq.(\ref{eq:3state})). 

Next, we analyze the time evolution of the initial state $\vert \psi\rangle=(1,0)^T\equiv\ket{+z}$, dependent on $A$, as expressed in Eq.(\ref{eq:errorprofile}). The observable $A$ can be decomposed as $A={1\!\!1} +\sigma_x$. Hence, the time evolution of the initial state yields 
\begin{eqnarray}
\vert\psi(t)\rangle&=&e^{- \mathrm{i} t A}\vert\psi\rangle=e^{- \mathrm{i} t ({1\!\!1} +\sigma_x)}\vert\psi\rangle\nonumber\\&\rightarrow&e^{- \mathrm{i} t \sigma_x}\vert\psi\rangle=e^{( \mathrm{i}\alpha  \sigma_x)/2}\vert\psi\rangle\nonumber\\&=&\Big({1\!\!1} \cos\frac{\alpha}{2}-\mathrm{i}\sigma_x\sin\frac{\alpha}{2}\Big)\ket{+z} \equiv \vert\psi(\alpha)\rangle,
\end{eqnarray}  
which is simply a rotation about the $x$-axis by an angle $\alpha$ (see Bloch sphere in Fig.\,\ref{fig:setup}). Thus the parametrization has changed from time $t$ to an (experimentally adjustable) spinor rotation angle $\alpha$.
\textcolor{black}{
\subsection{Sharp $M$ measurement}
In order to demonstrate the counter example a sharp measurement of $M$ is required. The decomposition of $M$ into projectors is denoted as 
\begin{eqnarray}  
M&=&\sqrt 2\,\,\Pi_1(\sqrt 2)-\sqrt 2\,\,\Pi_1(-\sqrt 2)\nonumber\\
&=&\sqrt 2\,\,{\rm P}^M(\sqrt 2)-\sqrt 2\,\,{\rm P}^M(-\sqrt 2),
\end{eqnarray}
where
\begin{eqnarray}  \label{eq:epsilonProj}
\hspace{-5mm} {\rm P}^M(\sqrt 2)&=&\frac{1}{2}\Big({1\!\!1} +\overbrace{\frac{\sigma_x+\sigma_z}{\sqrt 2}}^{\sigma_m}\Big)\nonumber\\
 {\rm P}^M(-\sqrt 2)&=&\frac{1}{2}\Big({1\!\!1} -\overbrace{\frac{\sigma_x+\sigma_z}{\sqrt 2}}^{\sigma_m}\Big),
\end{eqnarray}
with 
\begin{eqnarray}  
M^{(2)}=\sum_{x=\pm\sqrt{2}}x^{2}\, {\rm P}^M(x)=2\,{1\!\!1}=M^2. 
\end{eqnarray}
Therefore, the \emph{error-profile} $\varepsilon^2_\alpha(A,\Pi_1,\vert\psi(\alpha)\rangle)$ yields
\begin{eqnarray}\label{eq:5state}
&\varepsilon^2_\alpha(A,\Pi_1,\vert\psi(\alpha)\rangle) =\underbrace{\langle\psi(\alpha)\vert A^2\vert\psi(\alpha)\rangle}_{\neq 1}+\underbrace{\langle\psi(\alpha)\vert M^{2}\vert\psi(\alpha)\rangle}_{\neq 1}\nonumber\\
&-{\langle\psi(\alpha)\vert M\vert\psi(\alpha)\rangle}-\underbrace{\langle\psi(\alpha)\vert A\, M\,A\vert\psi(\alpha)\rangle}_{2\langle+x\vert M\vert+x\rangle}\nonumber\\
&+\overbrace{\langle\psi(\alpha)\vert \underbrace{(A-{1\!\!1})}_{\sigma_x}}^{\langle \psi(\alpha+\pi)\vert}\, M\,\overbrace{\underbrace{(A-{1\!\!1})}_{\sigma_x}\vert\psi(\alpha)\rangle}^{\vert\psi(\alpha+\pi)\rangle}\nonumber\\
&+\underbrace{ \langle\psi\vert M^{(2)}-M^2\vert\psi\rangle}_{=0},
\end{eqnarray}  
which finally gives
\begin{eqnarray} 
\varepsilon_\alpha (A,\Pi_1,\vert\psi(\alpha)\rangle) = 2\vert\sin \frac{\alpha}{2}\vert,
\end{eqnarray}
with \emph{locally uniform q-rms error}  $\bar\varepsilon (A,\Pi_1,\vert\psi\rangle) =2$, as predicted in \cite{Ozawa19}. Note that only for dichotomic measurements the first two terms of Eq.(\ref{eq:5state}) are unity and the error profiles becomes $\alpha$-independent (see Supplementary Information \cite{SuppPOVM} for experimental details and results of all individual expectation values of the \emph{sharp} $M$-measurement). }

\subsection{Unsharp $M$ measurement}
In addition, we performed generalized (unsharp) measurements, described by POVM $\Pi_2$, to determine the \emph{q-rms error profile} $\bar\varepsilon (A,\Pi_2,\vert\psi\rangle)$, where a decomposition of $M$ in terms of \emph{POVM elements} is applied, which is found as 
\begin{eqnarray} 
M=2\,\Pi_2(2)-2\,\Pi_2(-2).
\end{eqnarray}
The expectation value of $M$ is expressed as  
\begin{eqnarray} 
\langle  \psi(\alpha)\vert M \vert \psi(\alpha)\rangle=2p[\Pi_2(2),\psi(\alpha)]-2p[\Pi_2(-2),\psi(\alpha)],\nonumber\\
\end{eqnarray}
with probabilities $p[\Pi_2(2),\psi(\alpha)]=\mathrm{Tr}(\Pi_2(2)\,\rho_\alpha)$ and $p[\Pi_2(-2),\psi(\alpha)]=\mathrm{Tr}(\Pi_2(-2)\,\rho_\alpha)$, with $\rho_\alpha=\ketbra{\psi(\alpha)}$, being the probabilities of obtaining the respective results. The individual POVM elements are given by Eq.(\ref{eq:POVM}),
%\begin{eqnarray}  \label{eq:POVM}
%\Pi_2(2)&=&\frac{1}{2}\bigg( {1\!\!1} +\frac{1}{2} \sigma_x+\frac{1}{2} \sigma_z\bigg)\nonumber\\
%\Pi_2(-2)&=&\frac{1}{2}\bigg( {1\!\!1} -\frac{1}{2}\sigma_x-\frac{1}{2} \sigma_z\bigg),
%\end{eqnarray}
with $M^{(2)}=4\,{1\!\!1}\neq M^2=2\,{1\!\!1}$. This accounts for a \emph{generalized} measurement (with $\Pi_2(2)+\Pi_2(-2)={1\!\!1}$, obeying the completeness relation of POVMs).
%In Fig.\,\ref{fig:psiBloch}\,(c) a graphical illustration of the POVM elements can be seen.
Applying the definition of the \emph{q-rms error profile} $\varepsilon_\alpha$ from Eq.(\ref{eq:5state}) evidently reproduces the predictions for \emph{q-rms error profile} 
\begin{eqnarray} 
\varepsilon_\alpha (A,\Pi_2,\vert\psi(\alpha)\rangle) =\sqrt{4-2\cos\alpha},
\end{eqnarray}
and for the \emph{locally uniform q-rms error} we get $\bar\varepsilon (A,\Pi_2,\vert\psi\rangle) =\sqrt 6$. 

In the actual experiment the noisy POVM is realized by a \emph{randomized} combination of a projective measurement of $\sigma_m=\frac{1}{\sqrt 2}\sigma_z+\frac{1}{\sqrt 2}\sigma_x$ and a \emph{no-measurement}.
 %(see Fig.\,\ref{fig:psiBloch}\,(a)) and a \emph{no measurement}, as indicated by the unity operator in the POVM definitions from Eq.(\ref{eq:POVM}). 
The probability $p[\Pi_2,{\psi(\alpha)}]=\textrm{Tr}(\Pi_2\ketbra{\psi(\alpha)})$, is  measured by the projectors of $\sigma_m$, denoted as ${\rm{P}}^{\sigma_m}$, that is $\langle\psi(\alpha)\vert {\rm P}^{\sigma_m}\vert\psi(\alpha)\rangle$, together with a contribution of a \emph{no-measurement}. The \emph{no-measurement}, (\emph{identity}) is simply a measurement of spin operators, that are orthogonal to the plane spanned by the of the evolved states $\vert\psi(\alpha)\rangle$, namely  $\langle\psi(\alpha) \vert{\rm P}^{\sigma_x}(\pm 1) \vert\psi(\alpha)\rangle=\langle\psi(\alpha)\ketbra {\pm x} \vert\psi(\alpha)\rangle=\frac{1}{2}\,\,$ for all $\alpha\in [0,2\pi]$, and therefore add up to identity. 
We can thus rewrite the POVM elements as
\begin{eqnarray}  \label{eq:POVMRe}
&&\Pi_2(\pm 2)=\frac{1}{2}\bigg( {1\!\!1} \pm\frac{1}{2} \sigma_x\pm\frac{1}{2} \sigma_z\bigg)\nonumber\\
&=&\gamma_1 {1\!\!1} +\gamma_2\underbrace{ \frac{1}{2}\Big({1\!\!1} \pm\overbrace{\frac{\sigma_x+\sigma_z}{\sqrt 2}}^{\sigma_m}\Big)}_{\textrm{P}^{\sigma_m}(\pm 1)}\equiv\gamma_1 {1\!\!1} +\gamma_2\textrm{P}^{\sigma_m}(\pm 1)\nonumber\\
&=&\gamma_1\big(\textrm{P}^{\sigma_x} (1) +\textrm{P}^{\sigma_x} (-1)\big)+\gamma_2\textrm{P}^{\sigma_m}(\pm 1),
\end{eqnarray}
with $\gamma_1=\frac{1}{4}(2-\sqrt 2)$ as the \emph{weight} for the \emph{no-measurement} and $\gamma_2=\frac{1}{\sqrt 2}$ as \emph{weight} of the \emph{projector}. 
%The individual contributions can be seen Fig.\,\ref{fig:psiBloch}\,(b) for $\Pi^+_{m=+2}$ measurement.
Experimentally this is achieved, for example in the ${\rm{Tr}}(\Pi_2(2)\rho_\alpha)$ measurement, by controlling the current in DC coil 2 with a random generator, where with a frequency of 10\,Hz either the current $I^+_m$ for the ${\rm{P}}^{\sigma{_m}}(1)$ measurement or $I^\pm_{\rm{no}}$ for one of the two orthogonal spin components of the no-measurement is randomly chosen. The respective probabilities are given by  $p(I^+_{\rm{no}})=p(I^-_{\rm{no}})=\frac{1}{2}\frac{\gamma_1}{\gamma_1+\gamma_2}$ and $p(I_m)=\frac{\gamma_2}{\gamma_1+\gamma_2}$.

The same procedure is applied to the \red{measurement of} expectation values $\langle  \psi(\alpha)\vert (A-{1\!\!1})\,M\,(A-{1\!\!1}) \vert \psi(\alpha)\rangle$ and $\langle  \psi(\alpha)\vert A\,M\,A \vert \psi(\alpha)\rangle$. To obtain the results of $\langle  \psi(\alpha)\vert A^2 \vert \psi(\alpha)\rangle$ and $\langle  \psi(\alpha)\vert M^2 \vert \psi(\alpha)\rangle$ the expectation values of projector onto $\ket{ +x}$ and identity have to be measured (see Supplementary Information \cite{SuppPOVM} for experimental results of all individual expectation values of the \emph{unsharp} $M$-measurement).

\section{acknowledgements}

\vspace{-4mm}

\begin{acknowledgements}\textcolor{black}{
This work was supported by the Austrian science fund (FWF) Projects No. P 30677-N36 and P 27666-N20. M\,O acknowledges the support of the IRI-NU collaboration. \red{Y.H. is partly supported by KAKENHI.} }
\end{acknowledgements}

%\bibliographystyle{elsarticle-num} 
%\bibliography{!myBibliography}
%\bibliography{/Users/stephansponar/ownCloud/General/Tex/BibTex/!myBibliography}

%

\appendix
\onecolumngrid
 
 \noindent\rule{18cm}{0.4pt}
   
\vspace{6cm}

  \section{Supplementary Information}
 In this supplement, we provide details of the data evaluation, required for determination of \emph{error-profile}, accompanied by the underlying theoretical framework, completing the conceptual description given in the main text. On the next pages we present in detail  for the individual measurement results of all terms from Eq.(\ref{eq:3state})) for sharp (projective) measurements and unsharp (POVMs). 

 \renewcommand{\theequation}{S.\,\arabic{equation}}
  \renewcommand{\thefigure}{S.\,\arabic{figure}}
 \setcounter{equation}{0}
  \setcounter{figure}{0}

\vspace{-3mm}

\section{Experimental results of individual expectation values for projective measurements and POVMs.}

\vspace{-3mm}

\begin{itemize}
\item$\langle\psi(\alpha)\vert A^2\vert\psi(\alpha)\rangle:$
\end{itemize}

\begin{figure}[!h]
\begin{center}
	\includegraphics[width=0.49\textwidth]{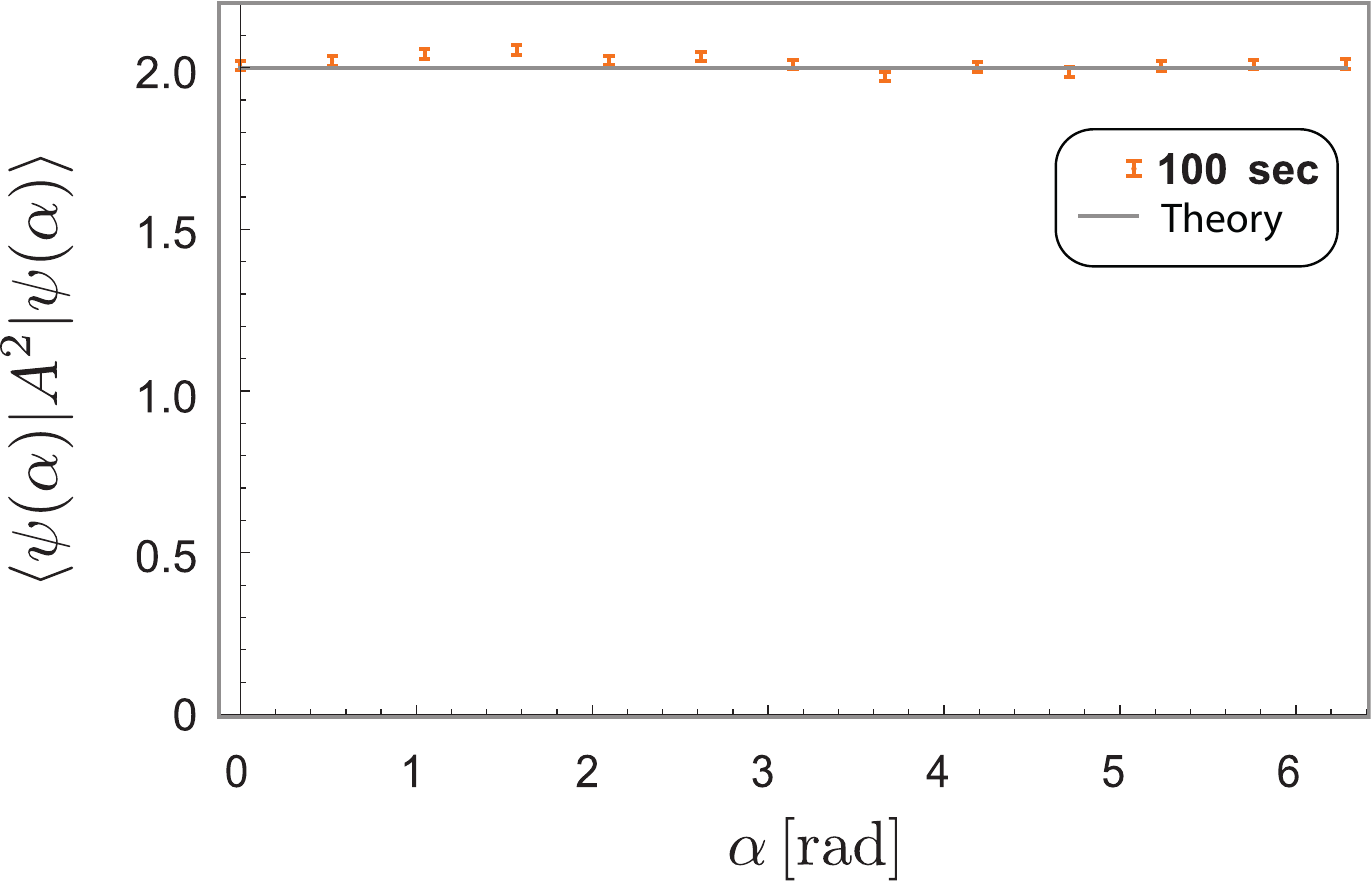}
	\caption{Experimental results of $\langle\psi(\alpha)\vert A^2\vert\psi(\alpha)\rangle=4\langle\psi(\alpha)\vert  {\rm P}^{\sigma_x}(1)\vert\psi(\alpha)\rangle=2$ for different evolved states $\vert\psi(\alpha)\rangle$ and measurements time $t_{\rm{maes}}=100\,$sec. (error-bars represent $\pm 1$ st. dev.).}
	\label{fig:ApI}
	\end{center}
\end{figure}

For the expectation value we can write $\langle\psi(\alpha)\vert A^2\vert\psi(\alpha)\rangle= \bra\psi (\alpha) (2\ketbra{+x})^2\ket{\psi(\alpha)} =\langle\psi(\alpha)\vert (2 {\rm P}^{\sigma_x}(1))^2\vert\psi(\alpha)\rangle=\langle\psi(\alpha)\vert 4(  {\rm P}^{\sigma_x}(1))^2\vert\psi(\alpha)\rangle=4\langle\psi(\alpha)\vert {\rm P}^{\sigma_x}(1)\vert\psi(\alpha)\rangle =4\frac{1}{2}=2$ for all values of $\alpha$. The experimental results of the \emph{first} of five expectation value from the three-state method from Eq.(\ref{eq:3state}) is depicted Fig.\,\ref{fig:ApI}.

\vspace{-3mm}

\begin{itemize}
\item$\langle\psi(\alpha)\vert M^2\vert\psi(\alpha)\rangle:$
\end{itemize}

Since $\langle\psi(\alpha)\vert M^{2}\vert\psi(\alpha)\rangle=\langle\psi(\alpha)\vert2\,{1\!\!1}\vert\psi(\alpha)\rangle=2\langle\psi(\alpha)\vert{1\!\!1}\vert\psi(\alpha)\rangle$, one has to measure the identity operator, by applying an appropriate decomposition, for instance expressed in form of $\langle\psi(\alpha)\vert{1\!\!1}\vert\psi(\alpha)\rangle=\langle\psi(\alpha)\vert(\ketbra{+x}+\ketbra{-x} )\vert \psi(\alpha)\rangle\\=\langle\psi(\alpha)\vert  {\rm{P}}^{\sigma_x}(1)+{\rm P}^{\sigma_x}(-1)\vert\psi(\alpha)\rangle=1$, with $(1/\sqrt{2},1/\sqrt{2})^T=\ket{ + x}$. The experimental results for the \emph{second} term from Eq.(\ref{eq:3state}) are depicted in  Fig.\,\ref{fig:MpI}.% A count rate based simulation is givenin Fig.\,\ref{fig:MpI}.
\begin{figure}[!h]
\begin{center}
	\includegraphics[width=0.49\textwidth]{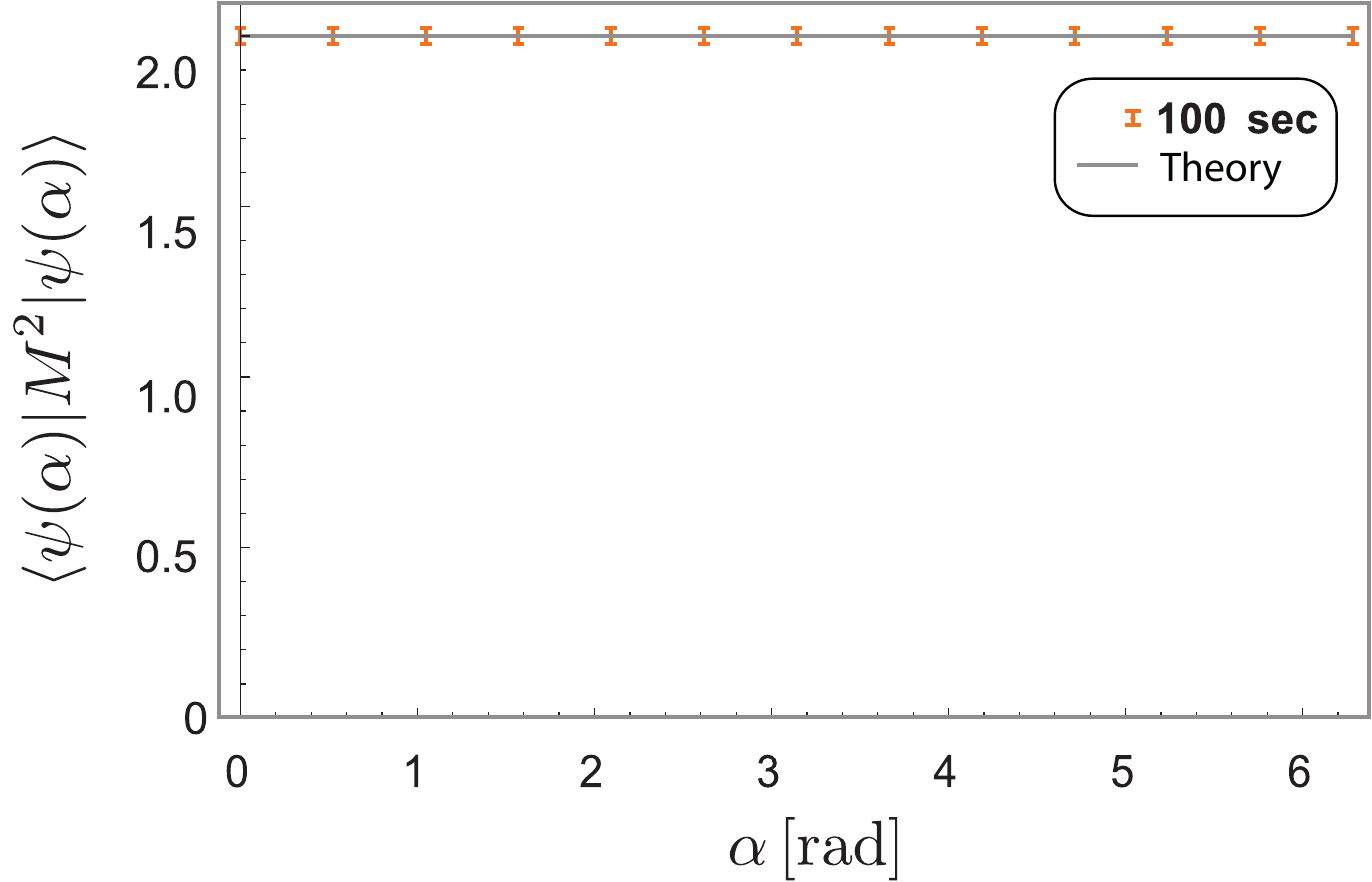}
	\caption{Experimental results of $\langle\psi(\alpha)\vert M^{2}\vert\psi(\alpha)\rangle=2$  for different evolved states $\vert\psi(\alpha)\rangle$ and measurements time $t_{\rm{maes}}=100\,$sec. (error-bars represent $\pm 1$ st. dev.).}
	\label{fig:MpI}
	\end{center}
\end{figure}

\newpage

\begin{itemize}
\item$\langle\psi(\alpha)\vert M\vert\psi(\alpha)\rangle:$
\end{itemize}
As discussed in the main text, using the POVM decomposition from Eq.(\ref{eq:POVM}), we can write 

\begin{eqnarray}
p[\Pi_2(2),\psi(\alpha)]=\mathrm{Tr}(\Pi_2(2)\,\rho_\alpha) =\gamma_1(\langle\psi(\alpha)\vert {\rm P}^{\sigma_x}(1) \vert \psi(\alpha)\rangle+\langle\psi(\alpha)\vert{\rm P}^{\sigma_x}(-1)\vert  \psi(\alpha)\rangle)+\gamma_2\,\langle\psi(\alpha)\vert {\rm P}^{\sigma_m}(1)\vert\psi(\alpha)\rangle,\nonumber\\
\end{eqnarray}
\begin{figure}[!h]
\begin{center}
	\includegraphics[width=0.9\textwidth]{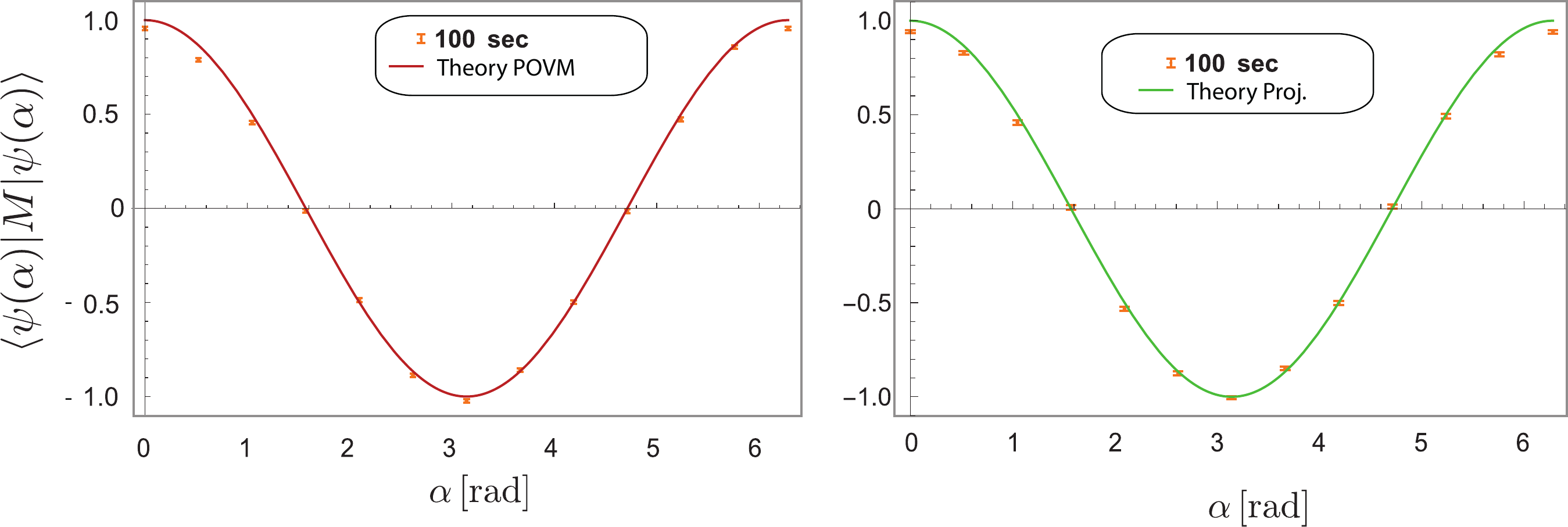}
	\caption{Experimental results of $\langle\psi(\alpha)\vert M \vert\psi(\alpha)\rangle$  for different evolved states $\vert\psi(\alpha)\rangle$ and measurements time $t_{\rm{maes}}=100\,$sec. (error-bars represent $\pm 1$ st. dev.).}
	\label{fig:M}
	\end{center}
\end{figure}

The final expectation value of $M$ in state $\vert\psi(\alpha)\rangle:$ can be expressed as 
\begin{eqnarray}
\langle\psi(\alpha)\vert M\vert\psi(\alpha)\rangle&=&2p[\Pi_2(2),\psi(\alpha)]-2p[\Pi_2(-2),\psi(\alpha)]\nonumber\\
&=&2\, \textrm{Tr}(\Pi_2(2)\ketbra{\psi(\alpha)}) -2\, \textrm{Tr}(\Pi_2(-2)\ketbra{\psi(\alpha)}\nonumber\\
&=&2\big(\gamma_1(\langle\psi(\alpha)\vert{\rm P}^{\sigma_x}(1)\vert \psi(\alpha)\rangle+\langle\psi(\alpha)\vert{\rm P}^{\sigma_x}(-1) \vert \psi(\alpha)\rangle)+\gamma_2\,\langle\psi(\alpha)\vert {\rm P}^{\sigma_m}(1)\vert\psi(\alpha)\rangle\big)\nonumber\\
&&-2\big(\gamma_1(\langle\psi(\alpha)\vert{\rm P}^{\sigma_x}(1)\vert \psi(\alpha)\rangle+\langle\psi(\alpha)\vert{\rm P}^{\sigma_x}(-1) \vert \psi(\alpha)\rangle)+\gamma_2\,\langle\psi(\alpha)\vert {\rm P}^{\sigma_m}(-1)\vert\psi(\alpha)\rangle\big).
\end{eqnarray}
For the projective (sharp) measurement we simply get 
\begin{eqnarray}
\langle\psi(\alpha)\vert M\vert\psi(\alpha)\rangle=\sqrt 2\langle\psi(\alpha)\vert  {\rm P}^{\sigma_m}(\sqrt 2)\,\vert\psi(\alpha)\rangle-\sqrt 2\langle\psi(\alpha)\vert  {\rm P}^{\sigma_m}(-\sqrt 2)\,\vert\psi(\alpha)\rangle,
\end{eqnarray}
with projectors ${\rm P}^\pm_m$ from Eq.(\ref{eq:epsilonProj}). 
The experimental results for the \emph{third} term from Eq.(\ref{eq:3state}) are depicted in  Fig.\,\ref{fig:M} (left) for generalized (unsharp) measurement via POVM decomposition, and (right) for a projective (sharp) measurement.

\begin{itemize}
\item$\langle\psi(\alpha)\vert A \,M\, A\vert\psi(\alpha)\rangle:$
\end{itemize}

\begin{figure}[!t]
\begin{center}
	\includegraphics[width=0.45\textwidth]{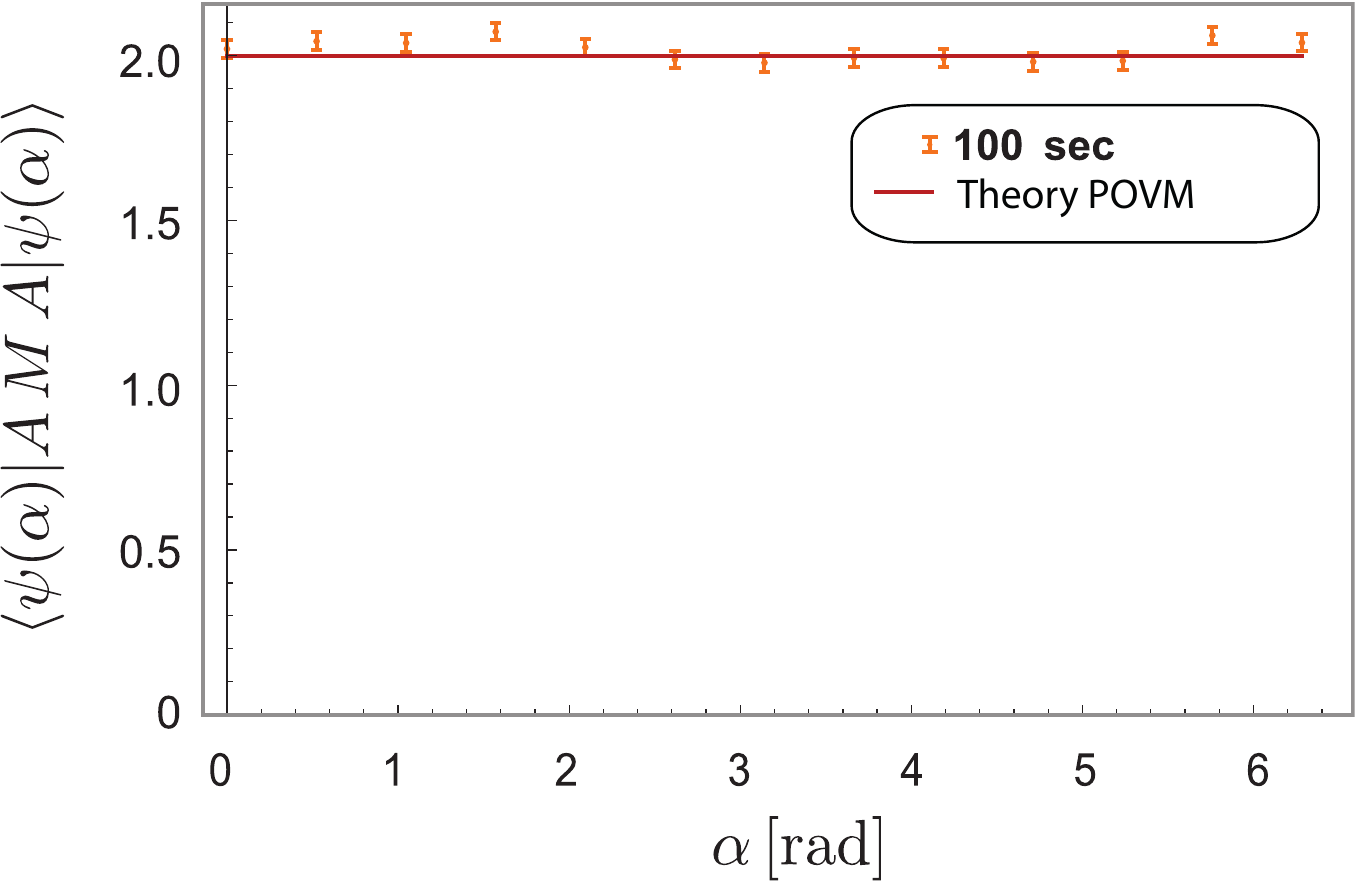}
	\caption{Experimental results of $\langle\psi(\alpha)\vert A\,M\, A\vert\psi(\alpha)\rangle=2$ for different evolved states $\vert\psi(\alpha)\rangle$ and measurements time $t_{\rm{maes}}=100\,$sec. (error-bars represent $\pm 1$ st. dev.).}
	\label{fig:AMA}
	\end{center}
\end{figure}

For the next expectation value $\langle\psi(\alpha)\vert A \,M\, A\vert\psi(\alpha)\rangle$ the procedure remains unchanged, however one has to take the \emph{effect} of $A$ on the evolved state into account, that is for example for the initial state 
\begin{eqnarray}
A.\vert\psi(\alpha=0)\rangle=(1,1)^T\equiv\vert\psi'(\alpha=0)\rangle.
\end{eqnarray}  
For example for $\alpha=\pi/2,$ we get 
\begin{eqnarray}
A.\vert\psi(\alpha=\pi/2)\rangle=\big((1-{\mathrm i})/\sqrt{2},(1-{\mathrm i})/\sqrt{2}\big)^T\equiv\vert\psi'(\alpha=\pi/2)\rangle.
\end{eqnarray}  
Although the state vector changes as a function of $\alpha$, the polarization vector $\vec P=\langle\psi(\alpha)\vert\hat\sigma\vert\psi(\alpha)\rangle$, with $\hat\sigma=(\sigma_x,\sigma_y,\sigma_z)$, remains completely unchanged namely $P=(2,0,0)^T$. So we only need to measure \emph{a single} expectation values, applying the procedure from the $M$ measurement, which accounts for all values of $\alpha$ and we get 
\begin{eqnarray}
\langle\psi(\alpha)\vert A \,M\, A\vert\psi(\alpha)\rangle=2\underbrace{\langle+x\vert  M \vert +x\rangle}_{=1}=2\underbrace{\big(2p[\Pi_2(2),+x]-2p[\Pi_2(-2),+x]\big)}_{=1}=2,
\end{eqnarray}  
or in more detail:
\begin{eqnarray}
\langle\psi(\alpha)\vert A\, M\, A\vert\psi(\alpha)\rangle&=&2\Big(2\, \textrm{Tr}[\Pi_2(2)\ketbra{+x}] -2\, \textrm{Tr}[\Pi_2(-2)\ketbra{+x}]\Big)\nonumber\\
&=&2\Big(2\big(\gamma_2\,\langle+x\vert {\rm P}^{\sigma_m}(1)\vert+x\rangle+\gamma_1(\langle+x\vert {\rm P}^{\sigma_z}(1)\vert +x\rangle)+(\langle+x\vert{\rm P}^{\sigma_z}(-1) \vert+x\rangle)\big)\nonumber\\
&&\quad-2\big(\gamma_2\,\langle+x\vert {\rm P}^{\sigma_m}(-1)\vert+x\rangle+\gamma_1(\langle+x\vert {\rm P}^{\sigma_z}(1)\vert +x\rangle)+(\langle+x\vert{\rm P}^{\sigma_z}(-1) \vert+x\rangle)\big)\Big)\nonumber\\
&=&2(2 \,\rm{x}\,0.75-2 \,\rm{x}\,0.25)=2.
\end{eqnarray}
For all values of $\alpha$, with $\gamma_1=\frac{1}{4}(2-\sqrt 2)$ as the \emph{weight} for the \emph{no-measurement} (now $\vert \pm z\rangle\langle \pm z\vert$) and $\gamma_2=\frac{1}{\sqrt 2}$ as \emph{weight} of the \emph{projector}.
Preparation of the state $\ket{+x}$ is experimentally achieved by rotating the initial state $\ket{+z}$ by $\pi/2$ (which results in a $\ket{+y}$ state) together with an additional displacement of DC-1 by a quarter of the Larmor period, that is  induced by the static magnetic guide field pointing in $+z$-direction.   
The experimental results for the \emph{fourth} term from Eq.(\ref{eq:3state}) are depicted in  Fig.\,\ref{fig:AMA} (left) for generalized (unsharp) measurement via POVM decomposition. 
Since also for projective (sharp) measurement the exprectation value is constant for all $\alpha$ and, unlike for POVMs, no randomness is involved the projective measurement is performed only once, resulting in $\langle\psi(\alpha)\vert A\, M\, A\vert\psi(\alpha)\rangle_{\rm{proj}}=2.04(2)$.

\begin{itemize}
\item$\langle\psi(\alpha)\vert  (A-{1\!\!1})\, M\,(A-{1\!\!1})\vert\psi(\alpha)\rangle:$
\end{itemize}

\begin{figure}[!b]
\begin{center}
	\includegraphics[width=0.88\textwidth]{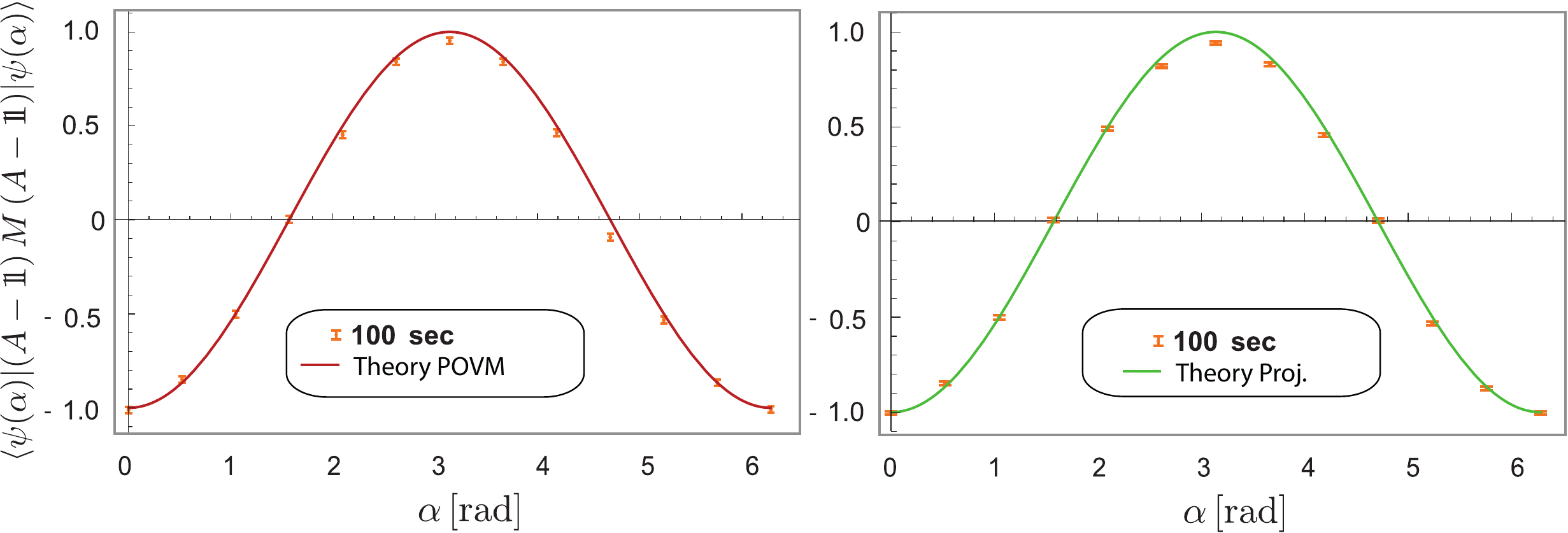}
	\caption{Experimental results of $\langle\psi(\alpha)\vert  (A-{1\!\!1})\, M\,(A-{1\!\!1})\vert\psi(\alpha)\rangle$ for different evolved states $\vert\psi(\alpha)\rangle$ and measurements time $t_{\rm{maes}}=100\,$sec. (error-bars represent $\pm 1$ st. dev.).}
	\label{fig:AIMIA}
	\end{center}
\end{figure}

For the fifth and last expectation value $\langle\psi\vert  (A-{1\!\!1})\, M\,(A-{1\!\!1})\vert\psi\rangle$ we get $A-{1\!\!1}=\sigma_x+{1\!\!1}-{1\!\!1}=\sigma_x$ which gives for the expectation value $\langle\psi(\alpha)\vert \sigma_x M\sigma_x\vert\psi(\alpha)\rangle= \langle\psi(\alpha+\pi)\vert  M\vert\psi(\alpha+\pi)\rangle$, hence a $M$ measurement as before, but with the orthogonal state of $\vert\psi(\alpha)\rangle$, denoted as $\vert\psi^\bot(\alpha)\rangle=\vert\psi(\alpha+\pi)\rangle$.
The final expectation value of $ (A-{1\!\!1})\, M\,(A-{1\!\!1})$ can be expressed in the POVM decomposition as 
\begin{eqnarray}
&&\langle\psi(\alpha+\pi)\vert M\vert\psi(\alpha+\pi)\rangle=2p[\Pi_2(2),\psi(\alpha+\pi)]-2p[\Pi_2(-2),\psi(\alpha+\pi)]\nonumber\\
&=&2\, \textrm{Tr}(\Pi_2(2)\ketbra{\psi(\alpha+\pi)}) -2\, \textrm{Tr}(\Pi_2(-2)\ketbra{\psi(\alpha+\pi)})\nonumber\\
&=&2\big(\gamma_1(\langle\psi(\alpha+\pi)\vert{\rm P}^{\sigma_x}(1)\vert \psi(\alpha+\pi)\rangle+\langle\psi(\alpha+\pi)\vert{\rm P}^{\sigma_x}(-1) \vert \psi(\alpha+\pi)\rangle)+\gamma_2\,\langle\psi(\alpha+\pi)\vert {\rm P}^{\sigma_m}(1)\vert\psi(\alpha+\pi)\rangle\big)\nonumber\\
&&-2\big(\gamma_1(\langle\psi(\alpha+\pi)\vert{\rm P}^{\sigma_x}(1) \vert\psi(\alpha+\pi)\rangle+\langle\psi(\alpha+\pi)\vert{\rm P}^{\sigma_x}(-1)  \vert\psi(\alpha+\pi)\rangle)+\gamma_2\,\langle\psi(\alpha+\pi)\vert {\rm P}^{\sigma_m}(-1)\vert\psi(\alpha+\pi)\rangle\big),\nonumber\\
\end{eqnarray}
with $\gamma_1=\frac{1}{4}(2-\sqrt 2)$ as the \emph{weight} for the \emph{no-measurement} (here again $\ket{ \pm x}\bra{ \pm x}$) and $\gamma_2=\frac{1}{\sqrt 2}$ as \emph{weight} of the \emph{projector}. The experimental results for the \emph{fifth} term from Eq.(\ref{eq:3state}) are depicted in Fig.\,\ref{fig:AIMIA} (left) for generalized (unsharp) measurement via POVM decomposition, and (right) for a projective (sharp) measurement.

\begin{itemize}
\item  $\langle\psi(\alpha)\vert M^{(2)}-M^2\vert\psi(\alpha)\rangle:$
\end{itemize}

Finally, in the case of unsharp (generalized) measurements also the non-zero term  $\langle\psi(\alpha)\vert M^{(2)}-M^2\vert\psi(\alpha)\rangle=\langle\psi(\alpha)\vert 4\,{1\!\!1}-2\,{1\!\!1}\vert\psi(\alpha)\rangle=2\langle\psi(\alpha)\vert {1\!\!1}\vert\psi(\alpha)\rangle$ has to be measured. The experimental results for the \emph{sixth} term from Eq.(\ref{eq:3state}) are depicted  in Fig.\,\ref{fig:MpI}.

\begin{figure}[!h]
\begin{center}
	\includegraphics[width=0.5\textwidth]{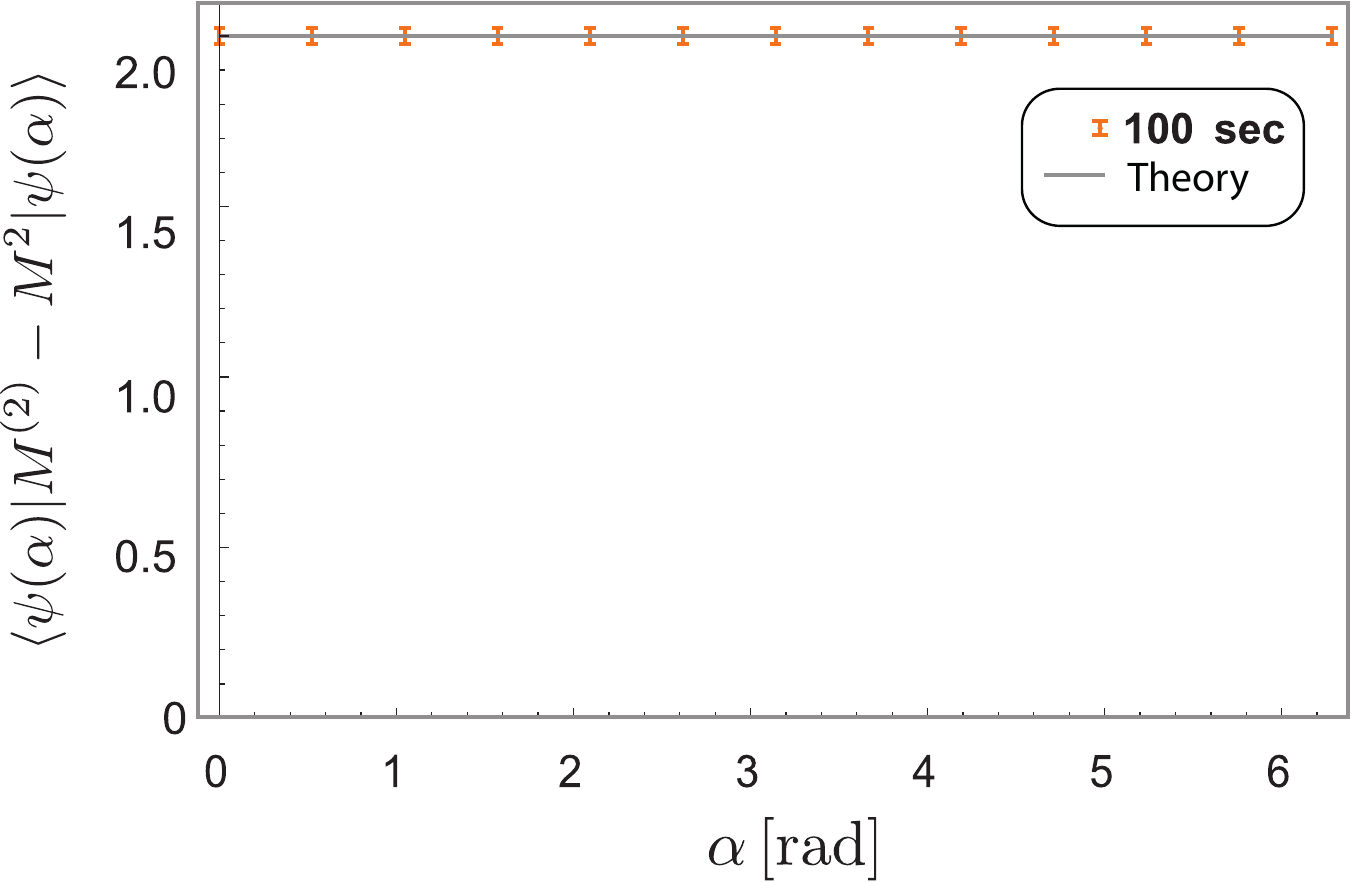}
	\caption{Experimental results of $\langle\psi(\alpha)\vert M^{(2)}-M^{2}\vert\psi(\alpha)\rangle=2$  for different evolved states $\vert\psi(\alpha)\rangle$ and measurements time $t_{\rm{maes}}=100\,$sec. (error-bars represent $\pm 1$ st. dev.).}
	\label{fig:MpI}
	\end{center}
\end{figure}

\end{document}